**Introduction:** *The Construction of Geo-Policy Barriers*

Perhaps the greatest development in rural communication technology in the middle of the 19th century was an eight foot board tied to the back of a horse. The King Road Drag scraped along country roads leveling humps and filling potholes.[1] The flat roads permitted rural postal deliveries, the development of parcel post and eventually a flood of information from catalog companies and publishers bearing goods and news from the big cities. Postal roads are two-way channels, of course, and although rural people could not take advantage of the commercial potential by mailing their livestock and produce, they could engage friends and relatives through letters as easily as order from Sears & Roebuck. Communication and community were encouraged by a combination of public policy, technology and commercial investment.

Geography conditions the ability of citizens to access to modern telecommunications networks as well, and to take advantage of communication technologies as a political, economic and cultural resource. In terms of the Internet, geographic factors comprise a kind of "capital." Similar to human capital or social capital, geographic capital can be most easily detected by its absence; the absence of geographic capital leads to an absence of functional Internet access.

Geo-policy barriers are chokepoints, mechanisms of control created through the interaction of geography, market forces and public policies. Geo-policy barriers not only constrict access but also construct both communication and communities, often in unintended ways. Geographic capital, therefore, can be impinged by three factors: the physical characteristics of places, the actions of telecommunications firms, and the actions of public policymakers.

---

[1] The importance of the King Road drag is illuminated in Gladwell, Malcolm. (1999, August). Clicks and Mortar, *New Yorker*.

The idea that infrastructure altered communication potential seemed obvious in the era of the King Road drag, when the words "transportation" and "communication" were virtually synonymous.[2] But what about today, at the dawn of the 21st century? In the era of the Internet, the ability of rural people to communicate, whether for purposes of commerce or community, is still shaped by public policy. At the federal level universal service policies have helped extend basic telecommunications to rural areas through a combination of regulatory mandate and wealth transfer. Rural telephone companies are crucial enterprises linking rural communities to the broader society and thus have been protected from many of the exigencies of the marketplace.[3] States have also created policies to assist rural citizens in obtaining basic telecommunications. States have used their role as arbiter of telephone tariffs to encourage the development of rural connectivity and have protected rural community phone companies from regulatory and market challenges through "rural exemptions."

Modern telecommunication systems in the U.S. are privately owned and managed, but have traditionally been highly regulated. The breakup of the regulated monopoly (AT&T) in 1984 and deregulatory efforts since 1996 have created a situation in which the marketplace plays a bigger role than at anytime in the last century.

---

[2] See Carey, James. (1989). Communication as Culture. Boston: Unwin Hyman.
[3] Section 3(37) of the Telecommunications Act of 1996, set forth in FCC Rules Part 51.5 designates rural telephone companies as meeting one of four criteria:
1. service area includes no incorporated area with more than 10,000 inhabitants or any territory in an urbanized area
2. the company provides telephone service to less than 50,000 access lines
3. the company provides telephone exchange service to any LEC with fewer than 100,000 access lines
4. the company has less than 15% of its access lines in communities of 50,000 or more inhabitants (as of the date of the Act, February 8, 1996.

Telecommunications firms have pledged, even demanded, to compete but little competition has come to residential customers so far. Even in areas where competition is beginning to emerge, the marketplace is highly structured by regulation. But these regulations seek only to mitigate market failure in remote areas, not to ensure adequate connectivity. The creation of exchange areas and Local Access Transport Areas (LATAs) reveals an underlying principle of centralization that corrals geographic capital within geo-policy barriers. This correspondence between market areas and regulatory boundaries has a couple of consequences for rural communities. First, it substitutes markets for communities, and assumes that the need to communicate is a function of dwelling within a contiguous market area. Second, geo-policy barriers can effectively discourage commercial investment in telecommunications infrastructure and impose additional burdens on citizens already struggling to connect to the Internet.

In the post-1996 telecommunications world, states are tasked with providing 75 percent of universal service.[4] Most states have developed a separate class of programs aimed at reducing the costs of long-distance communications for rural citizens. Commonly known as Expanded Area Service (EAS), these programs generally reduce intra-LATA long-distance costs either between specific exchanges or throughout a contiguous geographic area (Table 1). In the Internet era, the most important of these programs create flat-rate calling zones that allow remote customers to reach an Internet Service Provider (ISP) in a more populous area. By using EAS, remote customers can connect to the Internet from remote areas that do not support an ISP. Conversely, small ISPs can extend their markets in rural areas via EAS programs. But EAS programs do not always help where intended and can exacerbate the isolation and communication difficulties known as the "rural penalty."

The forces that shape basic telecommunications access – geography, commercial investment and public policy – will also determine who will bridge the digital divide in a broadband world. This study examines the how interplay of these forces structures communication access in rural Texas. No state has more people living in rural areas than Texas.[5] Perhaps no state so startlingly depicts the notion of a "digital divide." From "telecom alley" north of Dallas to the Silicon Hills of Austin and south to the international port of Houston, Texas is blessed with a robust network of high-speed communications technologies supporting some of the leading lights in telecommunications and computer industries. Away from this sliver of the state, however, the communications picture is less certain. This study compares findings from a statewide statistical analysis of rural Internet access with a case study of seven counties in west Texas, an area about the size of New England. The study examines role of the EAS program for rural Texans, known as Expanded Local Calling (ELC), in enabling access to the Internet. What emerges is a picture of a policy whose quirky contours and irregularities mimic the rugged geography it overlays.

**The Study:** *Expanded Local Calling in Rural Texas*

---

[4] For reviews of the share and administration of universal service programs, see Rosenberg, et. al, 1998 or the Benton Policy & Practice initiative, "The New Definition of Universal Service," online at: http://www.benton.org/Updates/summary.html#admin
[5] Texas has approximately 20 million residents spread over 261,000 square miles.

This study looks at the confluence of policy, geography and telecommunication provider investment in two ways: first, a statistical analysis of telephone exchanges is presented, then, a case study of seven rural counties illustrates some of the particular challenges facing remote citizens. (See Table 3 for study selection criteria).

The Texas version of ELC is similar to that of other states, although it varies in some particulars (Table 2). The distance, in air miles, between central telephone switches determines eligibility for the program. Rural residents can elect ELC if they are 22 or fewer miles from the exchange with which they want to connect.[6] Residents between 22 and 50 miles can also elect ELC, but they must prove a connection with the requested exchange, called "community of interest." Residents whose exchange is more than 50 miles from their nearest neighbor are ineligible, as are customers served by telephone coops and small telephone companies.[7]

There are 1,300 telephone exchanges in Texas of which about 208 represent primarily rural areas. Urban Influence Codes (UIC) determine "remoteness" by measuring county population and proximity to metropolitan areas.[8] This study incorporates UIC 7-9, counties with progressively small populations (under 10,000) non-adjacent to metropolitan counties. The primary interest of the study is the exchange level characteristics of remoteness, Internet access and public policy: the distance from the exchange to the nearest metropolis, the presence or absence of an ISP, and the presence or absence of ELC in the exchange. Remoteness was measured "as the crow flies" using a standard telephone exchange map. A survey of telephone exchanges conducted by the Texas Public Utility Commission in the spring of 2000 determined the presence or absence of ISPs in exchanges. Data from that survey were paired with data gleaned from an ELC database maintained by the Texas PUC to determine the impact of ELC policy on rural Internet access. Earlier studies, especially the NTIA "Falling Through the Gap" studies, have posited a role for demographic characteristics in shaping the digital divide NTIA, 2000). Many of those characteristics were collected at the county level for this study, using U.S. Census data to determine education level and Hispanic ethnicity levels.[9] The 1990 Census was also used to assign population and population density. The relative income of Texas counties is a state-level index that measures median income of families within a county as a percentage of median family income across all Texas counties.

---

[6] ELC elections proceed through ballots provided through phone company billing lists. Successful elections require a 70 percent majority of all customers who vote.
[7] Small companies in the case provide fewer than 10,000 customer lines. An exemption may also be made for any exchange with fewer than 10,000 access lines.
[8] Urban Influence Codes are available from the USDA online at: http://www.ers.usda.gov:80/briefing/rural/Data/

Telephone exchanges in Texas have both names, e.g., "Six Shooter," and numbers that correspond to the dialing prefix, or nxx code, for that area, e.g. "555-." Both names and numbers were used to ascertain the presence or absence of Expanded Local Calling in the 208 remote exchanges under study. When a caller in one exchange connects with another across the invisible exchange boundary, that call becomes subject to applicable non-local charges. Sometimes, the call is connected as a "local-long distance" toll call; sometimes it is connected at no additional charge. In the latter case, charges may be reduced as part of a commercial calling arrangement – a flat-rate offer made by the phone company – or as the result of public policy, such as Expanded Local Calling policies.

---

[9] Black and Asian populations were less than 2% in the areas studied. "Hispanic" is a complex identifier, particularly in a state that has been both a Spanish and a Mexican territory, and where Hispanics represent both the earliest and most recent non-indigenous inhabitants. The U.S. Census category "Hispanic" contains "white" and "non-White" sub-categories. This study uses the collective Hispanic category and does not distinguish between sub-categories.

As of spring 2000, more than 5,460 petitions for Expanded Local Calling had been filed with the Texas Public Utility Commission. Among those exchanges with rural Urban Influence Codes (UICs), 104 exchanges representing 49 Texas counties have made 725 ELC petitions, or about 13 percent of all ELC petitions.[10] About 30 percent, or 223, of those petitions were dismissed for some reason. That petition success rate holds for exchanges in each Urban Influence Code (6, 7 or 8), although exchanges with UIC 8 made more than half of all petitions. When we examine ELC success from the perspective of geographical areas, however, we can see that exchanges in some area codes are more successful than others. For instance, in our case study area, exchanges in the 830 area code (Central Texas area) made 23 petitions and were successful about 75 percent of the time. In the West Texas 915 area, residents made 160 petitions but more than 40 percent failed.

To provide some background for the case study, I will first discuss findings from the macro level analysis. Some of the reasons behind ELC success rates include the role of remoteness, the presence of exempt telephone carriers, and the possible role that social capital plays in matriculating through the ELC petition process.

---

[10] These figures are accurate as of spring 2000.

*Table 1: National Expanded Area Service Policy Variations*

| DIMENSION | ELIGIBILITY | MECHANISMS |
|---|---|---|
| COMMUNITY OF INTEREST | STATISTICAL | GEOGRAPHICAL |
| CUSTOMER SCOPE | INCLUSIVE | OPTIONAL |
| DIRECTIONALITY | UNI-DIRECTIONAL | BI-DIRECTIONAL |
|  |  |  |
| PRICING MECHANISM | FLAT | METERED |
| POLICY SCOPE | COMPREHENSIVE | CASE-BY-CASE |

*Table 2: Extended Local Calling Policy in Texas.*

| DIMENSION | MECHANISMS | RULE |
|---|---|---|
| COMMUNITY OF INTEREST | GEOGRAPHICAL | CO's 22-50 MILES APART |
| CUSTOMER SCOPE | INCLUSIVE | 70% OF THOSE RETURNING BALLOTS ELECT FOR ALL CUSTOMERS |
| DIRECTIONALITY | BI-DIRECTIONAL | ONLY ELECTING EXCHANGE CUSTOMERS PAY DIRECT FEES |
| PRICING MECHANISM | FLAT | $3.50/MONTH RESIDENTIAL $7/MONTH BUSINESS |
| POLICY SCOPE | COMPREHENSIVE | ALL RURAL AREAS SUBJECT TO PUC REGULATORY ACT |

*Table 3: Study Selection Criteria and Data Sources.*

| Level of Study | Study Area | Number of Cases | Area Selection | ISP Determination | ELC Determination | Demographic Data Source |
|---|---|---|---|---|---|---|
| **Case Study** | 5 West Texas 2 Central Texas counties | 25 telephone exchanges | Critical case and comparison group | ISP organization lists, screening phone calls, interviews | ELC petitions, ELC database, interviews | RUPRI exchange data |
| **Macro Level Study** | 49 rural counties throughout Texas | 208 telephone exchanges | Rural UIC codes (7, 8, 9) | PUC survey of rural telephone exchanges | ELC database | U.S. 1990 Census data |

**Expanded Local Calling and Internet Access**

The legacy of geo-policy barriers, such as exchange areas and LATAs, is mitigated by more recent policies, like ELC, when the policy corresponds to the needs of remote communities. Expanded Local Calling can help communities gain Internet access and expand their range of ISP options. However, ELC does not enhance prospects for connectivity in the most remote, sparsely populated areas. Distance restrictions and carrier exemptions appear to be most directly responsible for policy failure. A statistical analysis clarifies the complex relationship between ELC policy, remoteness and Internet connectivity.

No relationship in this study is as strong or consistent as that between Expanded Local Calling and Internet access (Table 4). The presence of ELC in an exchange is the key predictor of ISP presence (.507, $p<.000$). A rural exchange with ELC in place is twice as likely to have an ISP as those without the discount policy. This indicates that ELC can make a tremendous difference in the lives of those residents who can take advantage of it. By extending the reach of rural exchanges, ELC draws them closer into the sphere of advanced telecommunications networks and the benefits of the Internet. But nearly one-third of all ELC petitions are dismissed and we can presume that many other exchanges never attempt to gain the policy relief because they live in exempt areas. As it turns out, ELC is also negatively correlated to distance (-.210, $p<.003$), and positively correlated to population (.155, $p<.026$) and population density (.234, $p<.001$). These relationships allow us to understand Internet access in the context of ELC. The presence of Internet Service Providers is not correlated to population (-.007, $p<.919$), although it is somewhat correlated with population density (.115, p<.100). Thus, we are not seeing a phenomenon where ISPs simply locate in populated areas and ELC is coincidentally present. Rather it seems clear that expanded access to more metropolitan exchanges enables residents to take advantage of ISPs that may not necessarily be located in the local central office.

*Table 4: Variable Correlations*

**Correlations**

|  |  | ISP | ELC | DISTANCE | DENSTY | PCTHISP | EDU | INCOME | PCTUNEMP |
|---|---|---|---|---|---|---|---|---|---|
| ISP | Pearson Correlation | 1.000 | .507** | -.183** | .115 | -.206** | .072 | .198** | .085 |
|  | Sig. (2-tailed) | . | .000 | .009 | .100 | .003 | .304 | .004 | .224 |
|  | N | 207 | 207 | 204 | 207 | 207 | 207 | 207 | 207 |
| ELC | Pearson Correlation | .507** | 1.000 | -.210** | .234** | -.267** | -.096 | .006 | .106 |
|  | Sig. (2-tailed) | .000 | . | .003 | .001 | .000 | .170 | .932 | .128 |
|  | N | 207 | 207 | 204 | 207 | 207 | 207 | 207 | 207 |
| DISTANCE | Pearson Correlation | -.183** | -.210** | 1.000 | -.165* | .253** | .280** | -.194** | .132 |
|  | Sig. (2-tailed) | .009 | .003 | . | .018 | .000 | .000 | .006 | .060 |
|  | N | 204 | 204 | 204 | 204 | 204 | 204 | 204 | 204 |
| DENSTY | Pearson Correlation | .115 | .234** | -.165* | 1.000 | -.416** | -.302** | .070 | .316** |
|  | Sig. (2-tailed) | .100 | .001 | .018 | . | .000 | .000 | .319 | .000 |
|  | N | 207 | 207 | 204 | 207 | 207 | 207 | 207 | 207 |
| PCTHISP | Pearson Correlation | -.206** | -.267** | .253** | -.416** | 1.000 | -.218** | -.469** | .432** |
|  | Sig. (2-tailed) | .003 | .000 | .000 | .000 | . | .002 | .000 | .000 |
|  | N | 207 | 207 | 204 | 207 | 207 | 207 | 207 | 207 |
| EDU | Pearson Correlation | .072 | -.096 | .280** | -.302** | -.218** | 1.000 | .441** | -.424** |
|  | Sig. (2-tailed) | .304 | .170 | .000 | .000 | .002 | . | .000 | .000 |
|  | N | 207 | 207 | 204 | 207 | 207 | 207 | 207 | 207 |
| INCOME | Pearson Correlation | .198** | .006 | -.194** | .070 | -.469** | .441** | 1.000 | -.453** |
|  | Sig. (2-tailed) | .004 | .932 | .006 | .319 | .000 | .000 | . | .000 |
|  | N | 207 | 207 | 204 | 207 | 207 | 207 | 207 | 207 |
| PCTUNEM | Pearson Correlation | .085 | .106 | .132 | .316** | .432** | -.424** | -.453** | 1.000 |
|  | Sig. (2-tailed) | .224 | .128 | .060 | .000 | .000 | .000 | .000 | . |
|  | N | 207 | 207 | 204 | 207 | 207 | 207 | 207 | 207 |

**.Correlation is significant at the 0.01 level (2-tailed).

*.Correlation is significant at the 0.05 level (2-tailed).

**Table 5: Expanded Local Calling Correlations (Split Cases)**

ELC = 0            ELC Not in Exchange
ELC = 1            ELC Present in Exchange

**Correlations**

| ELC | | | ISP | DISTANCE | POP | DENSTY | PCTHISP |
|---|---|---|---|---|---|---|---|
| .00 | ISP | Pearson Correlation | 1.000 | -.041 | -.189 | -.075 | -.040 |
| | | Sig. (2-tailed) | . | .746 | .119 | .539 | .743 |
| | | N | 69 | 66 | 69 | 69 | 69 |
| | DISTANCE | Pearson Correlation | -.041 | 1.000 | -.101 | -.357** | .150 |
| | | Sig. (2-tailed) | .746 | . | .420 | .003 | .228 |
| | | N | 66 | 66 | 66 | 66 | 66 |
| | POP | Pearson Correlation | -.189 | -.101 | 1.000 | .685** | .039 |
| | | Sig. (2-tailed) | .119 | .420 | . | .000 | .748 |
| | | N | 69 | 66 | 69 | 69 | 69 |
| | DENSTY | Pearson Correlation | -.075 | -.357** | .685** | 1.000 | -.338** |
| | | Sig. (2-tailed) | .539 | .003 | .000 | . | .005 |
| | | N | 69 | 66 | 69 | 69 | 69 |
| | PCTHISP | Pearson Correlation | -.040 | .150 | .039 | -.338** | 1.000 |
| | | Sig. (2-tailed) | .743 | .228 | .748 | .005 | . |
| | | N | 69 | 66 | 69 | 69 | 69 |
| 1.00 | ISP | Pearson Correlation | 1.000 | -.117 | -.070 | .016 | -.102 |
| | | Sig. (2-tailed) | . | .172 | .412 | .856 | .232 |
| | | N | 138 | 138 | 138 | 138 | 138 |
| | DISTANCE | Pearson Correlation | -.117 | 1.000 | .117 | -.026 | .254** |
| | | Sig. (2-tailed) | .172 | . | .172 | .764 | .003 |
| | | N | 138 | 138 | 138 | 138 | 138 |
| | POP | Pearson Correlation | -.070 | .117 | 1.000 | .685** | .045 |
| | | Sig. (2-tailed) | .412 | .172 | . | .000 | .601 |
| | | N | 138 | 138 | 138 | 138 | 138 |
| | DENSTY | Pearson Correlation | .016 | -.026 | .685** | 1.000 | -.393** |
| | | Sig. (2-tailed) | .856 | .764 | .000 | . | .000 |
| | | N | 138 | 138 | 138 | 138 | 138 |
| | PCTHISP | Pearson Correlation | -.102 | .254** | .045 | -.393** | 1.000 |
| | | Sig. (2-tailed) | .232 | .003 | .601 | .000 | . |
| | | N | 138 | 138 | 138 | 138 | 138 |

**. Correlation is significant at the 0.01 level (2-tailed).

**ELC and Remoteness**

The relationship between distance, ELC and ISP presence is clearly illustrated by clustering the remoteness variable into categories (Table 5). We notice threshold effects for both ELC and ISP presence. Not only are remote exchanges less likely to have either ELC or an ISP[11] but the efficacy of ELC in enabling ISP presence also diminishes with distance because ELC does not apply after the 50 mile cutoff. For instance, 76 percent of all exchanges 50 or fewer miles from a metropolis have ELC and 72 percent have an ISP. Those exchanges in the middle range, between 51 and 76 miles out, 68 percent have ELC and less than half have an ISP. But for those between 75 and 100 miles from a metropolis, ELC rates are 50 percent while ISPs are present in just one third of exchanges. For the six exchanges over 150 miles, just two have ELC or an ISP. These figures suggest a couple of things. First, for exchanges between 50 and 75 miles from a metropolis -- the ideal group to be helped by existing ELC policy -- about two-thirds can take advantage of the policy and less than half have Internet access. Second, for exchanges more than 75 miles from a metropolis, ELC does not provide much help.

**ELC and Demographic Factors**

We might expect that ELC would be correlated to socio-economic variables. However, no significant relationship between ELC adoption and education or income emerged. Expanded Local Calling policy is most frequently exercised in those exchanges that lie in counties with higher and more concentrated populations, and where Hispanics comprise a relatively small portion of the populace (-.267, p<.000). It is difficult to analyze this relationship, particularly in light of the case study.[12] Hispanic concentrations are highest in remote counties with low population density, two factors that independently affect ELC presence. The role of ethnicity in adoption of advanced telecommunications has been noted in earlier studies; Hispanics nationally are less likely to own computers or access the Internet than other ethnic groups. But illuminating the precise interaction of ethnicity in ELC adoption will require further study.

**Case Study:** *Contrasting Expressions of Geography and Policy*

Differences within counties – in carriers, in exchange boundaries, and in distance – are revealed with a narrower focus on the exchange, rather than the county, as a unit of analysis. This study presents some of the challenges from legacy telecommunications policies and commercial investment patters facing rural residents. Of the 208 rural telephone exchanges in Texas, 25 lie in the general area of the case study.[13] Of these, 19 lie wholly within our seven county study area: nine exchanges in the five county west area, and ten exchanges in the two county central area. Because ELC is reciprocal in Texas (petitioned exchanges get the same dialing privileges as those that petition) the surrounding exchanges were examined to see if they had ELC connection into our study area.

---

[11] The exceptions are three remote exchanges clustered along a highway that runs through Alpine, Texas, the seat of a remote state university, served by the dominant carrier, Southwestern Bell.
[12] For instance Coyanosa (89% Hispanic) successfully petitioned for ELC and gained Internet access, whereas Valentine (16% Hispanic) failed in its attempt to boost connectivity.
[13] One of the key problems in analyzing telephone exchanges at the county level is that exchanges do not map precisely over counties and county level studies tend to mask geo-policy barriers. However,

**Central Texas Area:** *Proximity Aids in Overcoming Geo-Policy Barriers*

In sharp contrast to the West Texas study area, where short trips can take hours, many residents of Llano and Blanco counties drive an hour or less to tap the resources of the Austin and San Antonio metropolitan areas. Exchange areas in this region are smaller, and distances between central switches are generally shorter than their West Texas counterparts. The complicating factor in this area is not so much distance as federal regulatory boundaries that condition both telephone rates and ELC feasibility.

The Kingsland exchange in Llano County is the only successful Expanded Local Calling petitioner in our Central Texas study area. Kingsland customers are connected by ELC to three other exchanges – Marble Falls, Buchanan Dam, and Burnet – all of which lie outside Llano and Blanco counties. Kingsland petitioners overcame an additional hurdle, gaining federal approval to make toll-free inter-LATA connections in each case. Kingsland sits in the San Angelo SMA[14], whereas Burnet and Buchanan Dam are in the Austin LATA, and Marble Falls is in the San Antonio LATA. Inter-LATA discounts require FCC approval to circumvent the inter-exchange carrier requirement normally applicable in inter-LATA call transfers.

Although there was only a single successful case in the Central Texas area, it points to the role of public information in understanding public policy and to the confusing web of public and commercial responsibilities in telecommunications. Residents of Kingsland had the geographic good fortune and the community wherewithal to get ELC. Expanded Local Calling has also led directly to cheaper Internet access in at least one case in Central Texas. The Marble Falls ISP would incur higher costs without

---

demographic and connectivity data is generally collected at the county level, by both state and federal entities, making broad, generalizable studies at the exchange level difficult.

ELC. It operates in all three area LATAs in the area and avoids inter-LATA barriers by employing Remote Call Forwarding (RCF) devices. But RCFs can significantly impact line capacity, quality, and bandwidth reducing both Internet service quality and the serviceable number of Internet customers in a region.

ELC substitutes for remote forwarding devices in the Marble Falls network. For instance, the ISP employed an RCF in Kingsland serving Llano and Tow exchanges. Customers calling in from north and west of Kingsland were transferred via RCF to the Granite Shoals exchange, and then routed through to the ISP's headquarters in Marble Falls. With ELC, the Kingsland calls can be routed directly to Marble Falls, across the LATA, at the flat monthly ELC rate. Many ISPs consider the time and costs associated with maintaining equipment to be a significant growth inhibitor and a drain on cash flow; ELC saves the ISP costs associated with purchasing, operating and maintaining the remote call forwarding devices. Expanded Local Calling also expands the potential market of the Marble Falls ISP by accommodating a greater number of simultaneous connections and increasing line reliability. As broadband considerations come into play, ISPs with landline connections to their customers can employ a greater variety of broadband solutions over greater distances.

**West Texas Area**: *Small Distances Make the Difference in a Vast Area*

In an area equivalent in size to most of New England, only three of 17 exchanges – Coyanosa, Ft. Stockton, and Imperial– have implemented Expanded Local Calling.[15] Rural exchanges closest to metropolitan areas can use ELC to extend their local calling area, but remote exchanges are held in check by policy exemptions. All of the successful exchanges are located in Pecos County, and all of them are less than 100 miles from Midland.[16] Coyanosa extended its local calling area to Ft. Stockton, Grandfalls, Monahans and Pecos (in Reeves County) exchanges. Imperial elected Crane, Coyanosa, Ft. Stockton, Monahans and Odessa. Grandfalls, an exchange outside our specific study area, elected Imperial, giving Imperial reciprocal discounts. One result of these elections is to interconnect several exchanges in the northern section of the west Texas study area.

---

[14] Texas has two designated Service Management Areas, or SMAs, that function as LATAs.

In light of these ELC elections, two results stand out related to Internet access. Coyanosa, which is served by two local ISPs, can dial into Ft. Stockton, which has four (the net gain is only two ISPs since the two operating in Coyanosa also operate in Ft. Stockton). Imperial, served by a single local ISP, can dial into Ft. Stockton, the largest town in the study area, with four. Small net gains in total ISPs are not necessarily important; the ISP total in Ft. Stockton is only about half of what is considered a competitive market.[17] But in this case, customers do gain some significant advantages. First, the infrastructure of two Ft. Stockton ISPs can provide higher quality and higher speed service. These ISPs have access to T1 lines from Southwestern Bell and provide ISDN service to their Ft. Stockton customers.[18] Although ISDN is significantly impaired by distance from the central switch, the presence of this advanced service indicates a

---

[15] Coyanosa and Imperial completed the petition process and conducted successful elections; Ft. Stockton receives ELC dialing discounts via reciprocal privileges as a result of those elections.

[16] Ft. Stockton is about 90 miles from Midland; Imperial is 64 miles, and Coyanosa is approximately 77 miles.

[17] Greenstein, Shane. (1998). Universal Service in the Digital Age: The commercialization and geography of U.S. Internet access, National Bureau of Economic ResearchWorking Paper 6453, [online] Available at: www.nber.org.papers/w6453.

[18] A T1 uses two copper pairs to provide 1.54Mbps in optimal range. Integrated Services Digital Network (ISDN) is a series of standards designed to transmit over ordinary telephone copper twisted-pair wiring as

couple of things for rural customers. Better quality backhaul capacity from Ft. Stockton to Midland improves service reliability and throughput speeds for customers. Extending their reach to Ft. Stockton also puts these small towns within a market sphere that is receiving timelier infrastructure upgrades and has attracted the attention of nearby metropolitan service providers. Second, even the limited competition in the area appears to have an effect on prices. ISPs reported cutting prices to residential customers in light of competitor entry. One ISP cut prices from $30 to $19.95 per month. Third, ISPs operating out of Ft. Stockton are able to take advantage of special features from the dominant carrier, Southwestern Bell. One ISP obtained special prefix feature that allows customers in their service area to dial-up the ISP as a local call. Another ISP uses a "roll-over" feature, in which customers dial a local number that is then automatically forwarded to another number in a distant exchange. These features add incremental costs to ISP's operating expenses, but the customer experiences these features as seamless connections via a local telephone number

Remote residents gain access in one of two ways. For Imperial, the significant connectivity is gained through a two-step process. In the first step, remote customers, because of their relative proximity to Ft. Stockton, gain access to upgraded services and a relatively competitive market. Ft. Stockton is too far from Odessa or Midland to reach them via ELC.[19] However, because of its relatively large size and its position on a major highway in Southwestern Bell territory, Ft. Stockton it is located on the technological fringe of the Odessa metro area. It can make the second step to upgrades and competition available in a larger Internet market.

---

well as over more advanced infrastructure. Usually, ISDN offers one voice and two data channels, but can be split into as many as 30 channels. Transmission speeds range up to 128kbps.

Coyanosa residents access the benefits of a metropolitan market in a single step. By extending their local calling range to Odessa, Coyanosa, a high plains ranch town of fewer than 400 people bypassed by the major highways, becomes part of a 250,000 person telecommunications market! In Coyanosa the potential of ELC and of the Internet for rural areas can be fully realized. Telephone companies and other providers are more likely to upgrade infrastructure between Coyanosa and Odessa if customers, either ISPs or residents, demand it. ELC in this sense solves a "chicken and egg" problem relative to infrastructure and demand. It is difficult to assess demand for services in areas without significant exposure to the benefits of those services. ELC allows residents to experience a profusion of Internet provision options and services and provides a better view of the potential market for ISPs. Market cognizance may lead to greater marketing efforts, which in turn heighten awareness of Internet options for Coyanosa customers. Internet Service Providers looking to expand their market territory will look first to capture those customers who can reach them with a local call. ELC in Coyanosa effectively extends the potential reach of ISPs whose primary market is the Midland-Odessa metropolitan area.

When Coyanosa residents access national ISPs that serve Midland, they gain service portability, competitive pricing and new service options. They also gain access to faster backhaul infrastructure eliminating some of the problems that reduce Internet convenience and viability in remote areas. Access to improved, national services – the Internet as it was meant to be – introduces a new sense of connection in Coyanosa both with the Midland-Odessa metro and with the rest of the connected world. Interaction via telecommunications can function to reduce feelings of isolation for remote residents and has important reciprocal effects for businesses and institutions. Connectivity in Coyanosa can shrink the perceived distance between the small town and the metro as well as the actual costs of communication and service delivery. This perceptual shift "moves" Coyanosa closer to the metro or gives them the choice to move closer, extending the reach of small town businesses and consolidating the markets of metropolitan service and information providers.

If Coyanosa demonstrates the positive potential of Expanded Local Calling in West Texas, the Valentine exchange, in Jeff Davis County, demonstrates how the current ELC regulations can work to prevent connectivity. The Texas PUC dismissed all five of Valentine's petitions before election. Valentine failed to articulate a Community of Interest in three cases; one petitioned exchange lay outside the 50-mile limit, and Big Bend Telephone exercised its exemption as a cooperative in the fifth case. As a result, Valentine remains isolated with a single ISP and low-bandwidth service.

Any constituent wishing to petition for ELC receives some specific guidelines on ELC exemptions and community of interest criteria. How then is it that an exchange can petition five exchanges, two of which are exempted by statute and in three cases fail to articulate community of interest? The answer lies in the specific geographic and institutional situation of the town. Valentine is a remote town of 217 people, about half living under the poverty line. Located in Jeff Davis County, it is about 35 miles west of Ft. Davis (pop. 426), the county seat and about 50 miles north of Alpine, the largest town in the area and home of Sul Ross University. Because Valentine relies upon Big Bend Regional Medical Center in Alpine, residents must dial long distance for routine and emergency medical consultation.

Alpine was clearly the primary target of Valentine's ELC petition, but Valentine had what they felt was ample reason to bridge the boundaries of each of the five exchanges. Besides telephone exchange boundaries, another set of invisible lines segregates the people in the Valentine area. School districts do not match up with telephone exchanges or city limits. The Valentine Independent School District is the largest single employer in Valentine, with 18 employees, and some residents naturally work in neighboring districts. People living not far outside of town make long distance calls from home to work and work to home. Similarly, students in VISD turn to area ranches for employment and the district has created some school-to-work programs to help students get jobs nearby. But students who want to call from within the Valentine exchange to ranches as close as nine miles away must place long distance calls. Students commuting to Sul Ross in Alpine are unable to utilize the Internet access they pay for as part of their standard fees without placing a long distance call. Unpredictable and unexpected long distance charges are budget busters for poor American families (Horrigan, et al., 1995). In an area like Valentine, long distance charges may dissuade potential callers from checking in with their kids at home, calling their children's teachers, seeking employment, using an otherwise free Internet service, and even asking for routine medical advice.

Valentine made Community of Interest arguments like those above with a focus on Alpine. However, the Texas PUC judged Alpine to be 51.55 air miles from the Valentine exchange and thus beyond the 50 mile limit of ELC. Although Valentine citizens routinely drive to Alpine, it is unlikely any of them had measured the precise air miles between the switches, nor is that information available from phone companies.[20] Similarly, there is no particular reason Valentine residents should know that an independent telephone company serves the Alamito exchange. With those two exchanges ruled out, Valentine needed to make a Community of Interest case for the remaining three in Marfa, Ft. Davis and Van Horn. Their argument for local calling between adjoining school districts was judged insufficient.

A Valentine educator filed the petition, a typical situation in the ELC process. Schools are the hub of rural communities and rural educators are particularly attuned to the costs of parent-teacher communication among other things. Rural schools are generally a significant part of economic development plans, as well. In those areas where long-distance rates discourage communication, the school becomes a hub with no spokes connecting it to the encircling community. Coyanosa, Imperial and Valentine are very similar communities in many respects. Each town has fewer than 800 residents and similar demographics. A key difference between Coyanosa and Imperial – successful ELC petitioners – and Valentine appears to be distance from a metropolitan area. In Valentine, where using ELC to access a metropolitan area is out of the question, even the distance to the nearest large town proved to be unconquerable.

**Conclusions:** *Geo-Policy Barriers, Communication and Community*

Pathways of wire and glass are the 21$^{st}$ century postal roads and computer algorithms are smoothing the bumps and grooves. The Internet, however, has not yet managed to eliminate the rural penalty. Rather, development patterns are exacerbating the rural disadvantage. The problem is not simply one of infrastructure, but is created through the development of telecommunications markets and the legacy of federal and state communication policies.

---

[20] Point-to-point exchange distances can be obtained from the PUC through a formal Open Records request, but the most likely way to get it is through the ELC petition process.

The overlapping geographies of the information age constitute some of the greatest challenges to rural connectivity. Exchange areas, LATAs and other geo-policy boundaries arrange citizens into markets, rather than arranging firms to serve communities. Perhaps this should not be surprising since agencies are formed to regulate industries. However, the directive of the FCC is to "serve the *public* interest, convenience and necessity," and similar language can be found in the mission of most public utility agencies. Universal service programs and other regulatory relief, such as ELC, will work best when they place citizens and communities at the "center" and dissolve, rather than erect, barriers to interaction.

It is clear that EAS policies can help rural citizens extend their reach toward Internet access, and that rural Internet Service Providers can utilize the policy to dissolve market and technological barriers; however, the policies discriminate. ELC builds strict, arbitrary boundaries for "community." The most striking example of the Texas PUC's commitment to the firms they regulate is the exemption for rural telephone companies, an exemptions designed to ensure service to rural customers. This exemption creates a feudalistic correspondence between communities and monopoly providers wherein the presence of the small company dictates the fate of the citizen. This logic is extended by the Community of Interest rules that encourage citizens to conceptualize community according to the frequency and intention of telephonic communications with institutions, rather than with each other. Again, the ideology of regulation constructs communities as constituents of commercial and governmental institutions around which they revolve in close geographic patterns.

In terms of Internet access, the "rural penalty" can be usefully reconceptualized as a "remote penalty," with the most remote towns least likely to enjoy the fruits of the communication revolution. Ironically, these policies often exclude the very residents that stand to benefit most from their effective implementation, paring away remote communities through a series of exemptions and requirements that test the abilities of even trained policy experts. Meanwhile, requirements to form "communities of interest" border on Orwellian doublespeak as they suggest to citizens that community can be reconfigured as telephonic dependency, and the potential community available in the everyplace of cyberspace is blunted by artificial boundaries

**Sources**